\documentclass[referee]{chjaa}
\usepackage{graphicx,times}             
\newcommand{\Chandra}{{\it Chandra\/}}

\newcommand{\re}{$R_{\rm e}$}

\begin{document}

\title{X-Ray Properties of the Point Source Population in
the Spiral Galaxy NGC 5055 (M63) with {\it Chandra} }
\volnopage{Vol.0 (200x) No.0, 000--000}

\setcounter{page}{1}

\author{Bing Luo
  \inst{1}\mailto{}
\and Jiyao Chen
  \inst{1}
\and Zhongli Zhang
  \inst{2}
\and Yu Wang
  \inst{2}
\and Jingying Wang
  \inst{2}
\and Haiguang Xu
  \inst{2}}

 \institute{Department of Physics, Fudan University,
220 Handan Road, Shanghai 200433, PRC\\
\email{wenyu\_wang@sjtu.edu.cn}
\and Department of Physics, Shanghai Jiao Tong
University, 800 Dongchuan Road, Shanghai 200240, PRC\\
}

\abstract{
By analyzing the \Chandra\ ACIS S3 data we studied the X-ray properties of the low-mass
and high-mass X-ray binary populations in the nearby spiral galaxy NGC 5055. A total of
43 X-ray point sources were detected within the 2 effective radii, with 31 sources located
on the disk and the rest 12 sources in the bulge. The resolved point sources dominate
the total X-ray emission of the galaxy by accounting for about 80\% of the total counts in
0.3--10 keV. By carrying out the spectral fittings we calculated the 0.3--10.0 keV luminosities
of all the detected X-ray point sources and found that they span a wide range from a few
$10^{37}$ erg s$^{-1}$ to over
$10^{39}$ erg s$^{-1}$.
After compensating for the incompleteness at the low luminosity end, we find that the
corrected XLF of the bulge population is well fitted with a broken power-law model with a
break at $1.57^{+0.21}_{-0.20}\times 10^{38}$ erg s$^{-1}$,
while the profile of the disk population's XLF agrees with a single power-law
distribution with a slope of $0.93^{+0.07}_{-0.06}$. The disk population is significantly richer at
$^{>}_{\sim}2\times10^{38}$ erg s$^{-1}$
than the bulge population, inferring that the disk may have undergone relatively recent,
strong starbursts that significantly increased the HMXB population, although ongoing
starbursts are also observed in the nuclear region. Similar XLF profiles of the bulge and
disk populations were found in M81. However, in most other spiral galaxies different
patterns of the spatial variation of the XLF profiles from the bulge to the disk
have been observed, indicating that the star formation and evolution history may be more complex
than we have expected.
\keywords{galaxies: individual (NGC 5055)---X-ray: binaries---stars:
luminosity function---stars: formation}
}

\authorrunning{Luo et al.}
\titlerunning{X-ray Point Sources in NGC 5055 (M63)}
\maketitle

\section{Introduction}
The study of the X-ray properties of the bright point sources in
spiral galaxies, most of which are low-mass X-ray binaries (LMXBs)
and high-mass X-ray binaries (HMXBs), may provide us with valuable
observational constraints on the star formation and evolution history
in the disk and bulge of the host galaxy. With the superb high spatial
resolution of the \Chandra\ X-Ray Observatory, a large population of
X-ray point sources have been resolved for the first time in some
nearby spiral galaxies. In a few specific cases, distinct X-ray
characteristics have been revealed for the sources located in the
bulge and those in the disk, which can be interpreted as the
evidence for the spatial variation of stellar population composition
that reflects the differences in star formation history (e.g.,
Tennant et al. 2001; Kong et al. 2002; Soria \& Kong 2002). In this
paper we present a \Chandra\ study of the point sources in the
nearby starburst galaxy NGC 5055 (M63; SA(rs)bc), which is one
of the prototype Arm Class 3 flocculent galaxies (Elmegreen \&
Elmegreen 1987) that shows regular, two-arm spiral structure to
a radius of 4.0 kpc in the near-infrared band. The optical center
of the galaxy is identified at RA=13h15m49.25s Dec=+42d01m49.3s
(J2000; Maoz et al. 1996). The LINER nucleus is UV bright and is
surrounded by luminous young star clusters, showing clear stellar
absorption signatures (Maoz et al. 1998; Leitherer et al. 2002).
The inclination and position angles of the galaxy are deduced to
be $58^{\circ }$ and $103^{\circ}$, respectively
(Garcia-Gomez \& Athanassoula 1991). We organize
the paper as follows. In \S2, we describe the observation and
data reduction. In \S3, we present the imaging analysis. In \S4,
we investigate the X-ray properties of the detected point sources,
which includes the temporal and spectral analysis as well as the
calculations of hardness ratios and X-ray luminosity functions (XLFs).
Finally, we discuss and summarize the results in \S5 and \S6,
respectively. Throughout the paper, we quote errors at the 90\%
confidence level unless mentioned otherwise. We adopt a distance
of 7.2 Mpc to NGC 5055 (Michele \& Serra 1997), which is consistent with
the distance calculated from its redshift ($z=0.001681$) if
cosmological parameters $H_{0} = 70$ km s$^{-1}$ Mpc$^{-1}$,
$\Omega_{m} = 0.3$ and $\Omega_{\Lambda} = 0.7$ are adopted.
At this distance $1^{\prime}$ corresponds to about 2.1 kpc.

\section{Observation and Data Reduction}
NGC 5055 was observed with \Chandra\ with the CCD 0, 1, 2, 3, 6 and
7 of the \Chandra\ Advanced CCD Imaging Spectrometer (ACIS) in
operation on two separate occasions, which started on August 27,
2001 for a total exposure of 28.4 ks, and on April 15, 2001 for a
short duration of 2.4 ks, respectively. The events were telemetered
in faint mode and the data were collected with a frame time of 3.2
s. The CCD temperature was set at $-$120 $\textordmasculine$C. In
the August 27 observation from which the data of this work was
drawn, the center of galaxy was positioned on the ACIS S3 chip (CCD
7) with an offset of $31^{\prime\prime}$ from the nominal pointing for
the S3 chip, so nearly all the emission of the galaxy was covered by
the S3 chip. In the analysis that follows, we used the CIAO software
version 2.3 to process the data extracted from the S3 chip only. In
order to apply the latest calibration, we started with the Level-1
data. We kept events with ASCA grades 0, 2, 3, 4 and 6, and
excluded bad pixels, bad columns, and columns adjacent to bad
columns and node boundaries. In order to identify occasional periods
of high background, we extracted the lightcurve of the source-free
regions on the S3 chip in 2.5--7.0 keV where the background flares
are expected to be most visible. We found that up to about 5\% of
the total exposure time was affected by high background flares. By
excluding the contaminated intervals we obtained a clean exposure
of 27.0 ks for the analysis.

\section{X-ray Image}
In Figure 1a we plot the \Chandra\ S3 image of NGC 5055 in 0.3--10.0
keV in logarithmic scale. The image has been corrected for both
exposure and background, and has been smoothed by using a minimum
signal-to-noise ratio of 3 and a maximum signal-to-noise ratio of 5
per beam. We find that the X-ray emission from the galaxy is
dominated by a large population of X-ray point sources, which
contribute about 80\% of the total counts of the galaxy in
0.3--10.0 keV. The diffuse X-ray emission is nearly symmetric within
about $20^{\prime\prime}$, and is slightly elongated in east-west
direction in outer regions. No diffuse X-rays is significantly detected beyond
$\simeq 33^{\prime\prime}$, which is approximate the size of the bulge
($\simeq35^{\prime\prime}$; Baggett et al. 1998). We find that the
diffuse X-ray emission is peaked at RA=13h15m49.3s DEC=+42d01m45.5s
(J2000) where a bright nuclear X-ray point source is detected. The
position of the X-ray peak also coincides with the optical
and infrared centers of the galaxy to within $0.5^{\prime\prime}$.
In Figure 1b we show the optical image drawn from the Digital Sky
Survey (DSS) in linear scale, on which the locations of the 43
detected X-ray sources are marked with circles for comparison
(\S4.1).

\section{X-ray Point Sources}
\subsection{Detections}
We detected X-Ray point sources on the ACIS S3 image using the CIAO
tool celldetect with a signal-to-noise threshold of 3. We first
restricted the detections in 0.3--10.0 keV, and then crosschecked
the results in 0.5--7.0 keV. We also have crosschecked the results
by using the wavelet-based CIAO tool wavdetect and by
visional-examination. On the whole S3 CCD 46 sources are detected, of
which 43 sources are located within the 2 effective radii (1
effective radius = 1\re\ = $1.6^{\prime}$; Thornley 1996). In this
work we focus our study on these 43 sources only. The
spatial distribution of the 43 sources show a clear concentration
towards the galaxy center, inferring that most of them are
physically associated with the galaxy. Based on the results of the
deep \Chandra\ observations of the blank fields (Mushotzky et al.
2000), we performed Monte-Carlo simulations and found that only few
($^{<}_{\sim}5$) of the 43 point sources may be unrelated background
sources. By overlaying the positions of these sources on the DSS
image (Fig. 1b), we find that 31 of the 43 sources are detected
on the disk with a tendency to reside on the spiral arms. The
rest 12 sources are detected in the bulge, of which up to about $1$
source may be a disk population source that is misclassified into
the bulge population due to the project effect. At the X-ray peak a
nuclear source is detected, however it does not have the highest
count rate among the detected sources. We list the properties of
the 43 sources in Table 1, where we sort them in
the order of increasing projected distance from the center of the
galaxy.

\subsection{Temporal Variabilities}
After removing the intervals of strong background flares, we
extracted the 0.3--10.0 keV lightcurves of the detected sources that
each has more than 60 counts in total. We calculated the
Kolmogoroff-Smirov (K-S) statistic for each of the lightcurves
against the null hypothesis that the count rate of the source is
temporally invariant over the effective exposure time; if the source
is temporally invariant, the cumulative fraction of the count is a
diagonal from 0 to 1. We find that Src 22 (243 counts), Src 35 (600
counts) and Src 38 (102 counts) show significant evidence
for temporal variations on the 90\% confidence level (Fig. 3). This is not
likely to be caused by the variations of the local background, since
the K-S test gives negative results in the background variability.
For the central source (Src 1), the temporal variability is less
significant in terms of the K-S test, possibly because K-S tests are
most sensitive around the median value of the independent variable.
However, by calculating the variability parameter $S=(f_{\rm
max}-f_{\rm min})/\sqrt{\sigma_{f_{\rm max}}^{2}+\sigma_{f_{\rm
min}}^{2}}=1.1$, where $f_{\rm max}$ and $f_{\rm min}$ are the
maximum and minimum count rates, respectively, and $\sigma_{f_{\rm
max}}$ and $\sigma_{f_{\rm min}}$ are the corresponding errors, we find
that Src 1 is marginally variable on timescales of 1 $\sim$ 5 hr during which
its count rate changed by about $50\%$.

\subsection{Hardness Ratios}
Since the background-corrected count rates of most of the resolved
X-ray point sources are low, which makes it impossible to carry out
spectral analysis for each source, we turned to study the hardness
ratios of all the resolved sources that are defined as $H21=(M-S)/(M
+ S)$ and $H31=(H-S)/(H + S)$, where $S$, $M$ and $H$ are the
background-corrected counts extracted in 0.3--1.0 keV ($S$),
1.0--2.0 keV ($M$) and 2.0--10.0 keV ($H$), respectively. The same
approach has been adopted in earlier works (e.g., Sarazin et al.
2000). We list the calculated hardness ratios and the 1$\sigma$
errors in column 7 and 8 of Table 1, and plot $H31$ against $H21$ in
Figure 4. In the figure we show
the predicted hardness ratio distributions for an absorbed power-law
model by adopting column densities of $1.31\times10^{20}$ cm$^{-2}$
(the Galactic value; Dickey \& Lockman 1990) and $3.93\times10^{20}$
cm$^{-2}$, and photon indices of $\Gamma=0.0,~1.0,~2.0,~3.0$ and
4.0, as well as the hardness ratios for an absorbed blackbody model
by adopting the Galactic absorption and temperatures of $kT=0.5,~0.4,~0.3,~0.2$ and 0.1 keV.

We find that the distribution of the colors is similar to that of
M31 (Kong et al. 2002) and the nearly face-on, gas-rich spiral
galaxy M83 (Soria \& Wu 2003). Most of the sources lie in a broad
diagonal band extending from $(H21,H31)=(-1,-1)$ to (1,1). One
source (Src 19) that is located at a moderate distance
($1.1^{\prime}$) to the galaxy center has hardness ratios of about
$(-1,-1)$ and is thus identified as a supersoft source (SSS).
If the selection criterions of Supper et al. (1997) and Kahabka (1999)
are applied, 5 other sources (Src 6, 7, 9, 13 and 14) can be
identified as candidates of SSSs as well since they satisfy
$H31+\sigma_{H31}<-1$ and $H21<0$, or $H21+\sigma_{H21}<-0.8$.
If we slacken the selection limit to $H21<0.5$ and $H31<0.5$
(e.g. Swartz et al. 2002), Src 12 and 15 can also be classified as SSS
candidates. As is shown in Figure 4 the hardness ratios of these supersoft
sources can approximately be described with an absorbed black-body
model with the temperatures ranging from about 0.2 to 0.3 keV.
Src 31 has a hardness ratio of (1,1), which appears to have been
heavily absorbed by a column density significantly larger than the
Galactic value. Since Src 31 is located far away from the center of
the galaxy ($d>110^{\prime\prime}$) and its hardness ratios agree
with those expected by an absorbed power-law model with
an absorption of $\geq0.4\times10^{22}$ cm$^{-2}$ and a photon index of $\Gamma\sim0$, it is probably an
unrelated background AGN.

For comparison, within the 2\re\ region the measured mean hardness
ratios for the total X-ray emission of the galaxy are
$(H21,H31)=(-0.28,0.18)$. For the same region, the mean hardness
ratios are $(H21,H31)=(-0.14,0.13)$ for all the resolved X-ray point
sources and $(-0.46,0.23)$ for the unresolved diffuse emission.

\subsection{Spectral Analysis}
We first extracted and studied the cumulative spectrum of all the 42
off-center point sources resolved within 2\re. The backgrounds were
extracted from a carefully selected annulus around each source. By
applying the latest CALDB we have corrected for the charge transfer
inefficiency (CTI) and the continuous degradation in the ACIS
quantum efficiency, which is especially severe at lower energies. To
avoid the effects of calibration uncertainties at lower energies and
instrumental background at higher energies we restricted the
spectral analysis to the 0.7--7.0 keV energy band. We found that the
cumulative spectrum cannot be fitted with a single absorbed
power-law model, if the absorption is fixed at the Galactic value.
When the absorption was allowed to vary, however, the absorbed
power-law model can give an acceptable fit ($\chi^2/dof=152.7/116$).
 The resulting absorption and photon index are
$0.14\pm0.01\times 10^{22}$ cm$^{-2}$ and
$2.05\pm0.06$, respectively. With these parameters the total flux of all
the 42 off-center sources in 0.3--10.0 keV is calculated to
be $7.79 \times 10^{-13}$ erg s$^{-1}$ cm$^{-2}$. We also have
divided the resolved off-center sources into the disk and the bulge
populations and studied their cumulative spectra in 0.7--7.0 keV.
The cumulative spectrum of the disk
population can be marginally fitted with an absorbed power-law model
with an absorption of $0.27\pm0.03\times 10^{22}$ cm$^{-2}$ and a
photon index of $2.45\pm0.08$ ($\chi^2/dof=138.9/112$). The
spectrum of bulge population, on the other hand, can be well fitted
with a model that consists of a power-law component and a black body
component, both subjected to a common absorption that is consistent
with the Galactic value ($\chi^2/dof=60.1/55$). The obtained photon
index and temperature are $\Gamma=1.36\pm0.13$ and $kT=0.12\pm0.01$
keV, respectively.

By excluding all the detected point sources we examined the spectrum
of the diffuse emissions extracted in $<2$\re\ that includes the
contributions from both the unresolved point sources and the
inter-stellar medium (ISM). The background was extracted in a
source-free boundary region on the S3 chip as far away as possible
from the galaxy. We first attempted to fit the spectrum with an
absorbed power-law or an absorbed MEKAL model, but neither of them
gives an acceptable fit to the data, unless the absorption is allowed
to increase to unreasonable, physically meaningless values. Thus we
attempted to apply a model that consists of a MEKAL component to
represent the emission of the hot ISM, and a power-law component to
represent the contribution of X-ray binaries, with both components
subjected to a common absorption. Since the abundance of the MEKAL
component is poorly constrained, we tentatively fixed it to 0.1
$Z_\odot$, which is obtained by Tyler et al. (2004) for the central region of NGC 5055.
We found that when the absorption is fixed to the Galactic
value, the model provides us with an acceptable fit
($\chi^2/dof=63.4/51$), with a gas temperature of $0.30\pm0.02$
keV and a photon index of $2.47\pm0.26$ . The total
flux of the emission in 0.3--10.0 keV is $7.24 \times 10^{-13}$ erg
s$^{-1}$ cm$^{-2}$, of which about 54\% can be ascribed to the
power-law component.

Of the 43 X-ray sources resolved within 2\re, 10 sources (Src 1, 5,
8, 9, 16, 18, 22, 25, 27 and 35) have a total counts larger than
100. We extracted the individual spectra of these sources and the
corresponding background spectra in annular regions adjacent to where the
source spectra were extracted. We fitted each spectrum with an
absorbed power-law model and/or an absorbed multicolor disk
blackbody model (DBB), and list the results in Table 2.
The spectrum of the center source (Src 1) is well fitted by an absorbed power-law model with a
photon index of $1.61\pm0.15$, which is typical for those of
supermassive black holes in active galaxies. The 0.3--10.0 keV
luminosity of Src 1 corrected for the absorption is calculated
to be $3.16\times$$10^{38}$ erg s$^{-1}$, which is lower than that
of most low-luminosity AGNs. The spectrum of the brightest source
Src 35, which showed significant temporal variations during the
observation (\S4.2), can be fitted with an absorbed DBB model with
an inner disk temperature of $kT_{\rm in}=0.49\pm0.04$ keV. This
source is a candidate ultra-luminous X-ray source (ULX) since its
0.3--10.0 keV luminosity is $1.25\times10^{39}$ erg s$^{-1}$.

According to the previous studies of the SSSs in spiral galaxies (e.g.,
M101, Pence et al. 2001; M31, Kong et al. 2002; M83, Soria \& Wu 2003),
SSSs are probably white dwarfs fueled by accretion from their low-mass
companions. Indeed we find that the spectrum of the SSS candidate Src 9 can be
fitted with an absorbed DBB model with the inner temperature of
$0.15\pm0.01$ keV, which is consistent with that of an accreting white
dwarf.

\subsection{X-Ray Luminosity Functions of the Resolved Off-Center Point Sources}
\noindent{\it All Off-Center Sources Detected within 2\re}\\
\indent Assuming that all the off-center X-ray point sources
resolved within 2\re\ are located at the distance of NGC 5055, we
calculated their X-ray luminosities in 0.3--10.0 keV using the
best-fit spectral parameters for their cumulative spectrum. The
conversion factor of the counts is $2.13 \times 10^{36}$ erg
cts$^{-1}$, and the resulting luminosities range from $2.1 \times
10^{37}$ to $1.25 \times 10^{39}$ erg s$^{-1}$. With these we
construct the XLF and illustrate it in Figure 5a. Since the detection
of the point sources is not complete at the faint end of the
luminosity function, by adopting a method similar to that utilized
in, e.g., Kim and Fabbiano (2004) and Xu et al. (2005) we ran
Monte-Carlo simulations to create fake point sources on the S3 image
of NGC 5055 in 0.3--10.0 keV. In the simulations we created fake
sources by using the MARX package (Wise et al. 1997), and assumed
that the radial distribution of the fake sources at any given
luminosity follows the r$^{1/4}$ law (de Vaucouleurs 1948). At a
given luminosity, we determined how many of the fake
sources can be detected with the same technique used in \S4.1. In
such a way we corrected both the observed XLF and the background for the unresolved sources.
The corrected XLF are also shown in Figure 5b.

We fitted both the uncorrected and corrected cumulative XLFs with the software Sherpa by using either a
single power-law profile or a broken power-law profile
\begin{displaymath}
 N(>L)=N_0 \left\{ \begin{array}{ll}
                   (L/L_{38})^{-\alpha_l}& L<L_b\\
                   (L_b)^{\alpha_h-\alpha_l}(L/L_{38})^{-\alpha_h}& L>L_b
 \end{array} \right. ,\hspace{1em}
\end{displaymath}
where L/L$_{38}$ is the 0.3--10.0 keV luminosity in units of
$10^{38}$ erg s$^{-1}$, and $\alpha_l$ and $\alpha_h$ are the slope
indices for the lower and higher luminosity ends, respectively. We
find that for the uncorrected XLF, the fittings with the single
power-law model can be immediately rejected
($\chi^{2}/dof=150.2/21$). It overestimates the data at the high
luminosities and underestimates the data at the low luminosities. The
use of the broken power-law model, on the other hand, can
significantly improve the fittings and provide us with an acceptable
fit ($\chi^{2}/dof=15.8/19$). The best-fit parameters at the 90\% confidence level are
$L_{b}=2.53^{+0.56}_{-0.45} \times 10^{38}$ erg s$^{-1}$,
$\alpha_{l}=0.61^{+0.02}_{-0.03}$ and
$\alpha_{h}=1.96^{+0.58}_{-0.18}$. For the XLF corrected for the
effect of incompleteness at the faint end of the XLF,
the single power-law model also cannot give an acceptable
fit ($\chi^{2}/dof=125.8/22$), while the broken power-law model can
improve the fittings significantly ($\chi^2/dof=23.1/20$) with the
best-fit parameters $L_{b}=3.11^{+0.37}_{-0.48} \times 10^{38}$ erg
s$^{-1}$, $\alpha_{l}=0.71\pm0.05$ and
$\alpha_{h}=2.36^{+0.54}_{-0.43}$.\\
\\
\noindent{\it Bulge and Disk Populations}\\
\indent We studied the XLFs of the bulge and disk population
sources (Fig. 5) and found that the single
power-law model is inadequate to describe both the uncorrected XLFs of
the bulge population ($\chi^{2}/dof=68.9/9$) and the disk population
($\chi^{2}/dof=30.3/19$). The broken power-law model, however, gives
a good fit to the data with $L_{b}=1.68^{+0.12}_{-0.22}\times
10^{38}$ erg s$^{-1}$, $\alpha_{l}=0.31\pm0.12$ and
$\alpha_{h}=3.95^{+1.44}_{-1.19}$ for the bulge population
($\chi^2/dof=7.4/7$), and $L_{b}=3.11^{+0.37}_{-0.48}\times 10^{38}$
erg s$^{-1}$, $\alpha_{l}=0.71\pm0.05$ and
$\alpha_{h}=2.36^{+0.54}_{-0.43}$ for the disk population
($\chi^2/dof=23.1/20$). We then corrected the XLFs for the
incompleteness at the low energies using the method as is described
above for all the resolved off-center sources. For the bulge
population, a broken power-law is still needed to describe the XLF
with $L_{b}=1.57^{+0.21}_{-0.20}\times 10^{38}$ erg s$^{-1}$,
$\alpha_{l}=0.38^{+0.20}_{-0.23}$ and
$\alpha_{h}=2.28^{+1.30}_{-0.65}$ ($\chi^2/dof=7.4/7$). The
corrected XLF of the disk population is found to be nicely
consistent with a single power-law model with a slope of
$\alpha=0.93^{+0.07}_{-0.06}$ ($\chi^{2}/dof=16.3/19$).

We have crosschecked our results on the XLF profiles of all the resolved
off-center sources and disk population by excluding Src 35, the
brightest source that may bias the fittings. We found that within the
errors the best-fit parameters are consistent with those obtained
with Src 35 included.

\section{Discussion}
We calculated the 0.3--10.0 keV luminosities of 43 X-ray point sources
detected within 2\re\ of NGC 5055 and found that
they span a wide range from about $2.1\times10^{37}$ erg s$^{-1}$ to
$1.25\times10^{39}$ erg s$^{-1}$, which is typical for spiral galaxies
(M81; Tennant et al. 2001, NGC 1637; Immler et al. 2003)
and early-type galaxies (Xu et al. 2005 and references therein) at
similar distances. After compensating for the incompleteness at the
low luminosity end, we find that the corrected XLF of the bulge
population is well fitted with a broken power-law model, while the
profile of the disk population's XLF satisfies a single power-law
distribution. The disk population is significantly richer at the
high-luminosity end ($^{>}_{\sim}2\times10^{38}$ erg s$^{-1}$) than
the bulge population, inferring that the star formation history of
the bulge is distinct from that of the disk. In other words, the
disk may have undergone recent, strong starbursts that
significantly increased the HMXB population, although ongoing
starbursts are also observed in the nuclear region of NGC 5055. Quite
similar phenomenon has been found in M81 (also identified as a LINER
or Sy1.8) by Tennant et al. (2001), who reported that the XLF of the
bulge population exhibits a break at $\sim 4\times10^{37}$ erg
s$^{-1}$, and the XLF of the disk population can be fitted with a
single power-law.

Different patterns of the spatial variation of the XLF profiles
from the bulge to the disk are observed in other spiral galaxies. In
M83, a SAB(s)c galaxy that hosts a starburst nucleus (Soira \& Wu 2003),
the profile of the bulge XLF is consistent with a single
power-law with a slope of 0.8, while a break appears on the XLF of
the disk population at about $8\times10^{37}$ erg s$^{-1}$. This
might suggest that, unlike in NGC 5055 and M81, there is no
strong starbursts in M83 on the disk in the recent past, and the HMXBs dominate
the source population only in the nuclear starburst regions, rather
than over the entire disk. In M31 Primini et al. (1993), Shirey et
al. (2001), Kaaret (2002) and Kong et al. (2002) found that
the cumulative XLF of all the detected point sources has a distinct
break at $2-3\times10^{37}$ erg s$^{-1}$, a slope of $1.36\pm0.33$
for the high luminosity end and a slope of $0.47\pm0.09$ for the low
luminosity end. Kong et al. (2002) also studied the cumulative XLFs
of the point sources in the inner bulge, outer bulge and disk
regions. They found that there is a spatial variation
in the XLF profiles; both the break luminosity and the slope of the
cumulative XLF increase outwards monotonously from the inner bulge,
where the LMXBs dominate the X-ray emission of point sources, to the
disk. In a few other spiral galaxies where the XLFs have been
studied in detail, however, the report of a break on the XLF is not
available in literatures. These includes M101 (Pence et al. 2001), IC342
(Bauer et al. 2003; Kong 2003), NGC 891 (Temple et al. 2005),
NGC 1637 (Immler et al. 2003) and M51 (Terashima \& Wilson 2004), for
which the slopes of the XLFs range from about 0.5 to about 1.

The diversity of the XLF profiles of the X-ray sources in spiral
galaxies indicates that the star formation and evolution history may
be more complex than we have expected, so that further careful multi-band
investigations are needed. In addition, we suggest that, since typically
only 50-150 X-ray point sources are detected per galaxy, the uncertainty
introduced by the small number statistics may bias our conclusions on the XLF
profile to a certain extent. In Xu et al. (2005), by performing
Monte-Carlo simulations we showed that even if there is an universal break
on the XLFs of early-type
galaxies, the statistical errors definitely preclude us from
measuring it correctly. The same result can be applied to the
measurements of the XLF profiles of spiral galaxies.

\section{Summary}
We detected a total of 43 X-ray point sources (12 in the bulge and 31 on the disk) within
the 2 effective radii of NGC 5055, whose 0.3--10.0 keV luminosities range from a few
$10^{37}$ erg s$^{-1}$ to over
$10^{39}$ erg s$^{-1}$.
After compensating for the incompleteness at the low luminosity end, the corrected XLF
of the bulge population is well fitted with a broken power-law model ($L_{b}=1.57^{+0.21}_{-0.20}\times 10^{38}$ erg s$^{-1}$),
while the XLF of the disk population satisfies a single power-law distribution ($\alpha=0.93^{+0.07}_{-0.06}$).
The disk population is significantly richer at
$^{>}_{\sim}2\times10^{38}$ erg s$^{-1}$
than the bulge population, inferring that the disk may have undergone relatively recent,
strong starbursts that significantly increased the HMXB population, although ongoing
starbursts are also observed in the nuclear region. This is similar to what has been
found in M81. In most other spiral galaxies, however, different patterns of the spatial
variation of the XLF profiles from the bulge to the disk are observed, indicating that
the star formation and evolution history may be more complex than we have expected.

\begin{acknowledgements}
This work was supported by the National Science Foundation of China (Grant No. 10273009 and 10233040),
Shanghai Key Projects in Basic Research No. 04JC14079.
\end{acknowledgements}


\begin{table}
\renewcommand{\thefootnote}{\alph {footnote}}
\caption{X-ray Properties of the Detected Sources $^{\rm a}$  }

\begin{center}\begin{tabular}{ l c c c c c r r| l}
\hline\hline\noalign{\smallskip}
 Src.  & R.A.(J2000) & Dec.(2000)  &count rate    & d        &
$L_X$~(0.3--10.0 keV)      &
 $H_{21}$
$^{\rm b}$
 & $H_{31}$
$^{\rm c}$
 & notes
 $^{\rm d}$\\
 No.& (h:m:s)     & (d:m:s)      &($10^{-3}$ cnt s$^{-1}$)&(arcsec)
   &($10^{38}$ erg s$^{-1}$)&  &  &\\
\hline\noalign{\smallskip}
1  & 13:15:49.287 & +42:01:45.41 &6.70$\pm0.51$ & 0         &  3.83           &$ 0.17\pm0.04$ &$-0.14\pm0.04$      & b, c, v  \\
2  & 13:15:49.423 & +42:01:46.23 &2.88$\pm0.34$ & 1.727     &  1.44           &$ 0.51\pm0.11$ &$ 0.87\pm0.03$        & b,    \\
3  & 13:15:50.109 & +42:01:38.90 &0.53$\pm0.14$ & 11.23     &  0.30           &$ 0.45\pm0.17$ &$ 0.65\pm0.11$         &  b,    \\
4  & 13:15:48.581 & +42:01:35.48 &0.46$\pm0.13$ & 12.66     &  0.24           &$ 0.66\pm0.16$ &$ 0.74\pm0.12$         & b      \\
5  & 13:15:49.570 & +42:01:27.06 &4.51$\pm0.40$ & 18.61     &  2.56           &$ 0.76\pm0.05$ &$ 0.86\pm0.03$        & b       \\
6  & 13:15:50.061 & +42:01:26.12 &0.99$\pm0.19$ & 21.12     &  0.58           &$-0.56\pm0.08$ &$-1\pm0.01$       & b, s       \\
7  & 13:15:48.033 & +42:01:28.42 &0.56$\pm0.15$ & 21.99     &  0.32           &$-0.77\pm0.08$ &$-1\pm0.01$       & b, s       \\
8  & 13:15:50.923 & +42:02:00.13 &3.76$\pm0.37$ & 23.42     &  2.16           &$ 0.25\pm0.06$ &$ 0.06\pm0.06$   & b         \\
9  & 13:15:51.380 & +42:01:40.38 &7.51$\pm0.52$ & 23.86     &  4.16           &$-0.89\pm0.02$ &$-1\pm0.01$      & b, s, v   \\
10 & 13:15:47.807 & +42:01:24.85 &3.19$\pm0.34$ & 26.35     &  1.79           &$ 0.30\pm0.07$ &$ 0.51\pm0.05$        & b       \\
11 & 13:15:50.042 & +42:02:15.21 &2.80$\pm0.32$ & 30.96     &  1.60           &$ 0.28\pm0.06$ &$-0.02\pm0.07$     & b       \\
12 & 13:15:46.642 & +42:02:00.89 &3.37$\pm0.36$ & 33.28     &  1.92           &$-0.63\pm0.04$ &$-0.98\pm0.01$      & b, s, v \\
13 & 13:15:52.813 & +42:01:47.54 &0.67$\pm0.15$ & 39.35     &  0.37           &$-0.27\pm0.11$ &$-1\pm0.01$        & s   \\
14 & 13:15:52.013 & +42:01:13.47 &1.05$\pm0.20$ & 44.07     &  0.61           &$-0.55\pm0.08$ &$-1\pm0.01$        & s  \\
15 & 13:15:45.619 & +42:01:17.90 &0.39$\pm0.12$ & 49.26     &  0.21           &$-0.59\pm0.13$ &$-0.77\pm0.09$      & s    \\
16 & 13:15:53.232 & +42:02:14.79 &8.87$\pm0.57$ & 52.87     &  5.10           &$ 0.10\pm0.03$ &$-0.31\pm0.03$     &           \\
17 & 13:15:44.417 & +42:02:20.17 &0.58$\pm0.15$ & 64.43     &  0.34           &$-0.17\pm0.13$ &$-0.47\pm0.11$     &         \\
18 & 13:15:53.889 & +42:01:04.68 &4.83$\pm0.42$ & 65.49     &  2.80           &$ 0.62\pm0.04$ &$ 0.58\pm0.05$       &          \\
19 & 13:15:43.364 & +42:01:50.30 &0.73$\pm0.16$ & 66.17     &  0.42           &$-1\pm0.01$    &$-1\pm0.01$     & s        \\
20 & 13:15:55.369 & +42:02:11.11 &0.84$\pm0.17$ & 72.47     &  0.46           &$ 0.66\pm0.11$ &$ 0.70\pm0.09$       &       \\
21 & 13:15:56.358 & +42:01:35.35 &1.42$\pm0.23$ & 79.43     &  0.83           &$ 0.33\pm0.09$ &$ 0.33\pm0.09$       &    \\
22 & 13:15:40.858 & +42:01:49.16 &8.71$\pm0.56$ & 93.98     &  5.02           &$ 0.16\pm0.03$ &$-0.22\pm0.04$      & v         \\
23 & 13:15:43.900 & +42:00:32.75 &0.55$\pm0.14$ & 94.25     &  0.31           &$ 0.86\pm0.08$ &$ 0.73\pm0.14$         &         \\
24 & 13:15:42.048 & +42:00:46.67 &0.68$\pm0.15$ & 99.78     &  0.37           &$ 0.05\pm0.12$ &$-0.62\pm0.10$     &          \\
25 & 13:15:39.338 & +42:01:53.27 &5.89$\pm0.46$ & 111.1     &  3.42           &$ 0.66\pm0.03$ &$ 0.43\pm0.05$      &           \\
26 & 13:15:57.370 & +42:00:39.78 &1.02$\pm0.21$ & 111.4     &  0.65           &$-0.14\pm0.10$ &$-0.19\pm0.09$  &               \\
27 & 13:15:39.739 & +42:01:11.00 &5.79$\pm0.46$ & 111.8     &  3.36           &$ 0.24\pm0.04$ &$-0.33\pm0.04$      &            \\
28 & 13:15:58.286 & +42:00:55.74 &0.87$\pm0.18$ & 111.9     &  0.50           &$ 0.49\pm0.11$ &$ 0.46\pm0.11$        &    \\
29 & 13:15:49.488 & +41:59:51.84 &2.16$\pm0.18$ & 113.5     &  1.25           &$ 0.36\pm0.07$ &$ 0.05\pm0.08$        &     \\
30 & 13:15:39.496 & +42:02:26.35 &2.43$\pm0.30$ & 116.5     &  1.41           &$ 0.75\pm0.05$ &$ 0.74\pm0.05$        &  \\
31 & 13:15:59.716 & +42:01:25.17 &0.45$\pm0.13$ & 117.9     &  0.26           &$ 1\pm0.01$    &$ 1\pm0.01$          &       \\
32 & 13:15:37.668 & +42:02:12.76 &1.46$\pm0.23$ & 132.3     &  0.84           &$ 0.24\pm0.09$ &$ 0.02\pm0.09$  &            \\
33 & 13:16:00.013 & +42:02:45.36 &0.82$\pm0.17$ & 133.6     &  0.45           &$ 0.05\pm0.11$ &$-0.68\pm0.09$  &               \\
34 & 13:16:00.916 & +42:00:50.72 &0.46$\pm0.13$ & 140.6     &  0.26           &$ 0.80\pm0.10$ &$ 0.63\pm0.17$        &   \\
35 & 13:16:02.251 & +42:01:53.63 &21.4$\pm0.88$ & 144.6     &  12.49          &$ 0.48\pm0.02$ &$-0.19\pm0.03$  & v             \\
36 & 13:15:36.348 & +42:01:24.17 &1.32$\pm0.22$ & 145.7     &  0.75           &$ 0.53\pm0.10$ &$ 0.66\pm0.07$         &    \\
37 & 13:15:55.308 & +42:04:00.39 &1.29$\pm0.22$ & 150.7     &  0.74           &$ 0.22\pm0.09$ &$-0.15\pm0.10$   &               \\
38 & 13:15:43.878 & +41:59:10.46 &3.50$\pm0.35$ & 166.2     &  2.02           &$ 0.01\pm0.05$ &$-0.43\pm0.05$      & v, u       \\
39 & 13:15:40.339 & +42:03:59.03 &0.74$\pm0.16$ & 166.6     &  0.44           &$ 0.59\pm0.10$ &$ 0.39\pm0.13$        &        \\
40 & 13:15:46.141 & +42:04:28.37 &0.77$\pm0.17$ & 166.6     &  0.45           &$ 0.32\pm0.11$ &$-0.10\pm0.13$      &           \\
41 & 13:15:48.505 & +42:04:32.15 &1.01$\pm0.19$ & 166.9     &  0.59           &$ 0.34\pm0.11$ &$ 0.28\pm0.11$    &            \\
42 & 13:15:37.359 & +42:03:32.19 &1.21$\pm0.21$ & 170.4     &  0.71           &$ 0.54\pm0.08$ &$ 0\pm0.12$          &      \\
43 & 13:15:34.029 & +42:02:26.31 &0.61$\pm0.15$ & 174.8     &  0.29           &$ 0.48\pm0.15$ &$ 0.53\pm0.14$      &             \\
\noalign{\smallskip}\hline
\end{tabular}\end{center}
{\footnotesize
$^{\rm a}$~The columns are arranged as follows: (1) source number;
(2)-(3): right ascension and declination (J2000); (4) count rate and
its error; (5) projected distance {\it d} to the center of the
galaxy; (6) intrinsic X-ray luminosity $L_{X}$, assuming that the source
is located at the distance of NGC 5055 and only subjected to the
Galactic absorption ($1.31 \times 10^{20}$ cm$^{-2}$; Dickey \&
Lockman 1990); (7)-(8): hardness ratios and the 1$\sigma$ errors (\S4.3); and (9) notes.}\\
\\
{\footnotesize $^{\rm b,~c}~H_{21}$ = (M-S)/(M+S), $H_{31}$ = (H-S)/(H+S)} \\
\\
{\footnotesize $^{\rm d}$~Abbreviations are b: source in bulge; c: central source;
s: candidate of supersoft source; u: ultra luminous x-ray source;
and v: source shows temporal variability. }

\end{table}

\begin{table}
\renewcommand{\thefootnote}{\alph {footnote}}
\caption{Spectral Fittings of the Brightest Sources$~^{\rm a}$ }
\begin{center}\begin{tabular}{lcccccccccc}
\hline\hline\noalign{\smallskip}
&\multicolumn{4}{c}{Power-law Model}&&\multicolumn{4}{c}{DBB Model} \\
\cline{2-5} \cline{7-10}\\
 Src.   & $N_H$               & $\alpha$     & $\chi^{2}/dof$   &$F_X$/$10^{-14}$                  && $N_{\rm H}$                 & $T_{\rm in}$      & $\chi^{2}/dof$   & $F_X$/$10^{-14}$       \\
  No.     & ($10^{22}$ cm$^{-2}$)&              &                  &(erg s$^{-1}$ cm$^{-2}$)  &&($10^{22}$ cm$^{-2}$)  & (keV)              &                  &(erg s$^{-1}$ cm$^{-2}$)
 \\
 \hline\noalign{\smallskip}
 1     &0.013/fixed          &$1.61\pm0.15$ & 7.7/7            &5.10                              &&0.013/fixed            &$0.94\pm0.13$  &3.5/7            &3.62 \\
 5     &0.013/fixed          &$0.23\pm0.14$ & 3.5/4            &7.64                              &&       $-$             & $-$           & $-$             & $-$ \\
 8     &0.013/fixed          &$1.41\pm0.16$ & 1.2/3            &3.95                              &&0.013/fixed            &$1.23\pm0.22$  &1.9/3            &2.67\\
 9     &  $-$      &  $-$     &  $-$                            &  $-$                            &&0.013/fixed            &$0.15\pm0.01$  &5.0/5            &4.29\\
 16    &0.013/fixed          &$1.67\pm0.12$ & 7.8/9            &7.18                              &&0.013/fixed            &$0.96\pm0.10$  &7.9/9   & 5.18\\
 18    &0.013/fixed          &$0.15\pm0.27$ & 1.4/3            &17.14                             &&$0.20\pm0.15$          &$3.02\pm7.34$  &0.39/2  &7.48         \\
 22    &0.013/fixed          &$1.46\pm0.13$ & 10.6/9           &7.85                              &&0.013/fixed            &$1.01\pm0.12$  &8.5/10           &5.00\\
 25    &$0.60\pm0.05$        &$2.34\pm0.33$ & 1.5/5            &5.02                              &&$0.30\pm0.10$          &$1.05\pm0.17$ &2.9/5            &4.34      \\
 27    &$0.37\pm0.17$        &$2.90\pm0.88$ & 4.9/4            &3.11                              &&$0.19\pm0.10$          &$0.49\pm0.12$  &4.9/4            &2.30\\
 35    &$0.63\pm0.09$        &$3.74\pm0.34$ & 19.6/22          &10.75                             &&$0.33\pm0.05$          &$0.49\pm0.04$  &17.4/22          &8.96\\
\hline\noalign{\smallskip}
\end{tabular}\end{center}
{\footnotesize $^{\rm a}$~Errors are quoted at the 90\% confidence level.}
\end{table}

\begin{figure}
\centering
\includegraphics[width=8cm,angle=0]{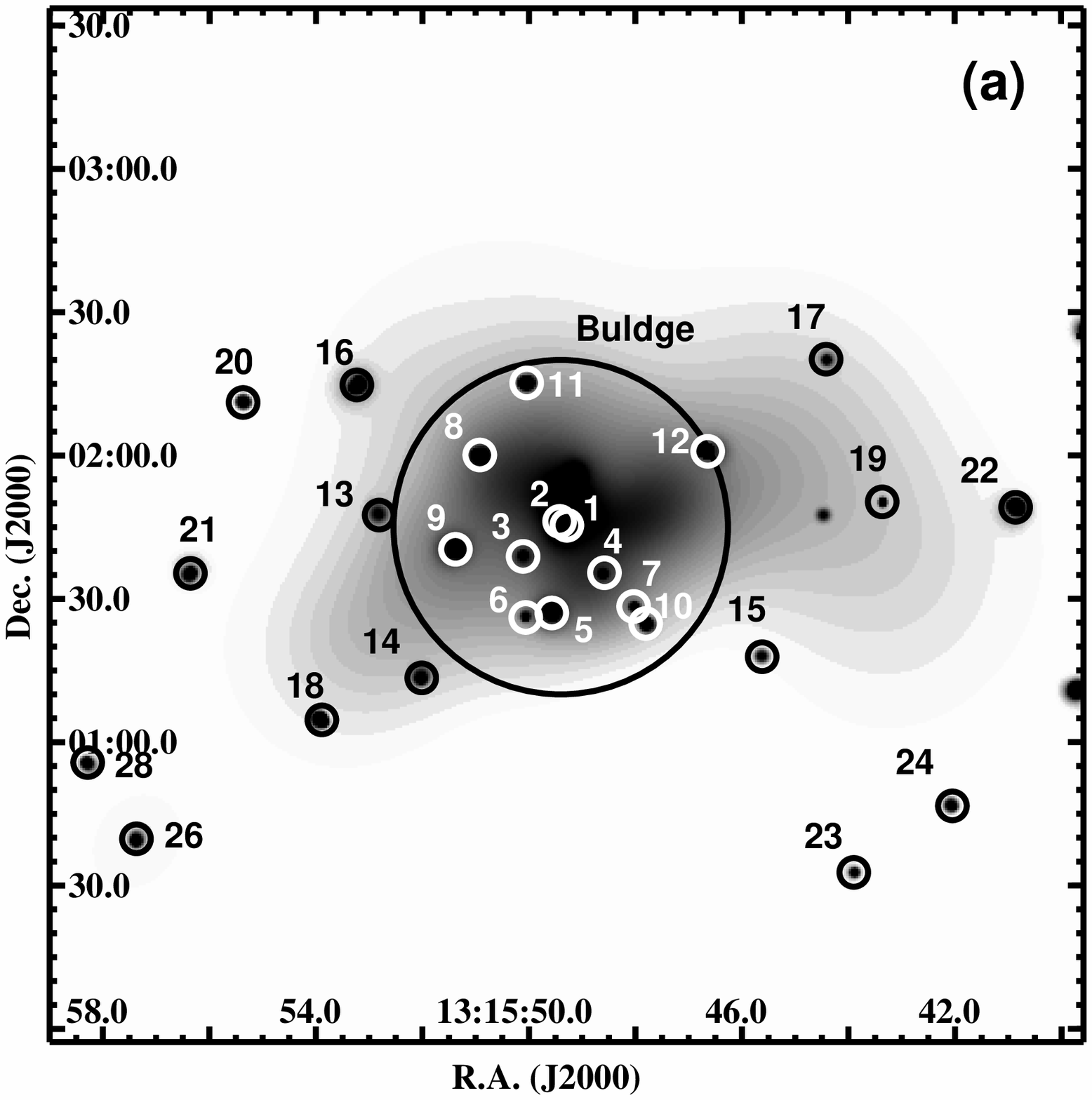}\includegraphics[width=8cm,angle=0]{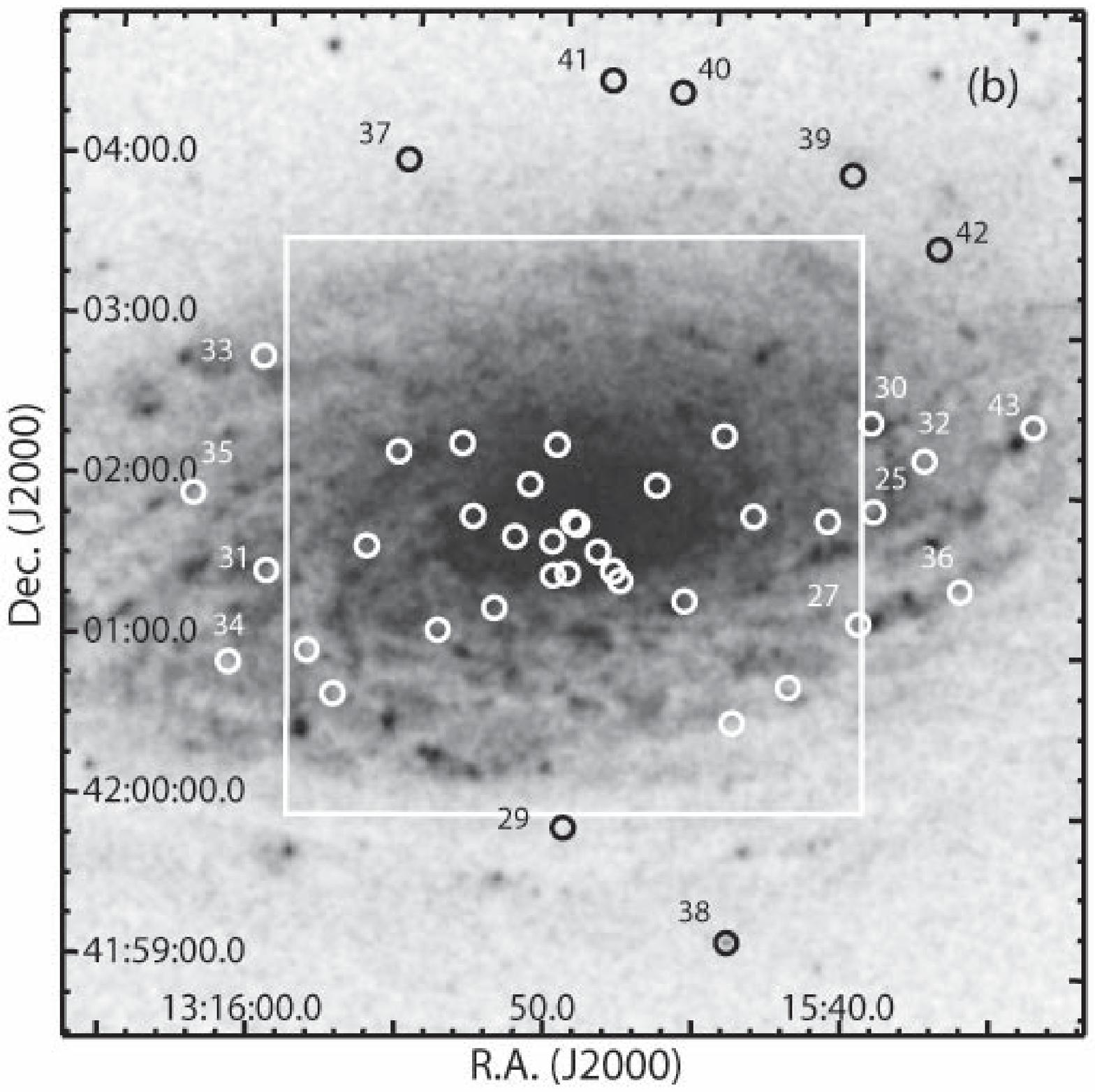}
\caption{(a): X-ray image of the central $3.6^{\prime}\times 3.6^\prime$
region of NGC 5055 in 0.3--10 keV in logarithmic scale, which has been smoothed
with a minimum significance of 3 and a maximum significance of 5, and has been
corrected for both exposure and background. (b): DSS blue image with the field of view of (a) and
the bulge marked with a box and a circle, respectively. On both images
the detected X-ray point sources are marked with small circles.
\label{fig:Fig1.ps}}
\end{figure}

\begin{figure}
\centering
\includegraphics[width=8cm,angle=270]{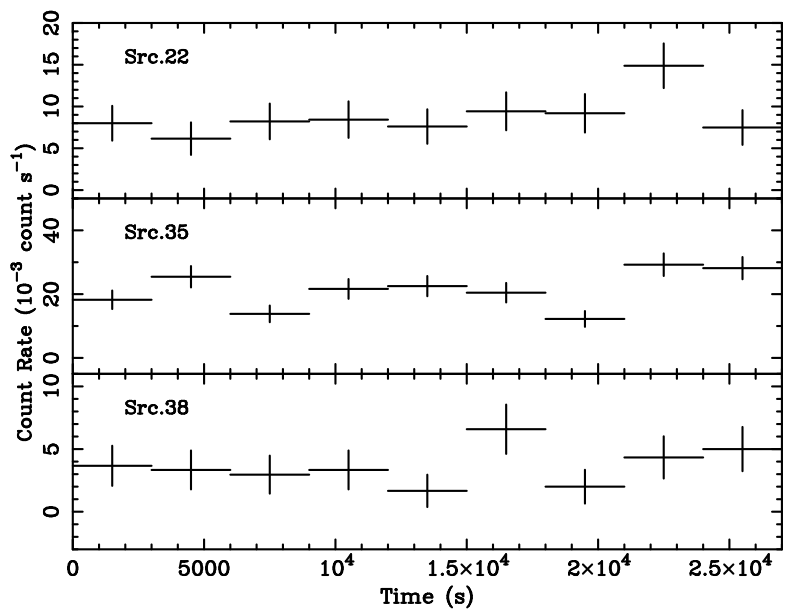}
\caption{Background-corrected lightcurves of Src 22, Src 35 and Src 38.}
\end{figure}

\begin{figure}
\centering
\includegraphics[width=8cm,angle=270]{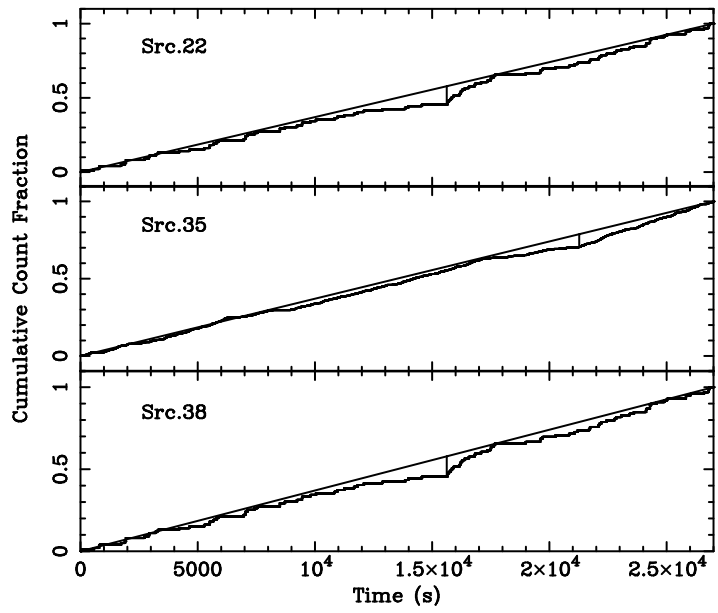}
\caption{Results of Kolmogoroff-Smirov test for the lightcurves of Src 22, Src 35 and Src 38.}
\end{figure}

\begin{figure}
\centering
\includegraphics[width=8cm,angle=270]{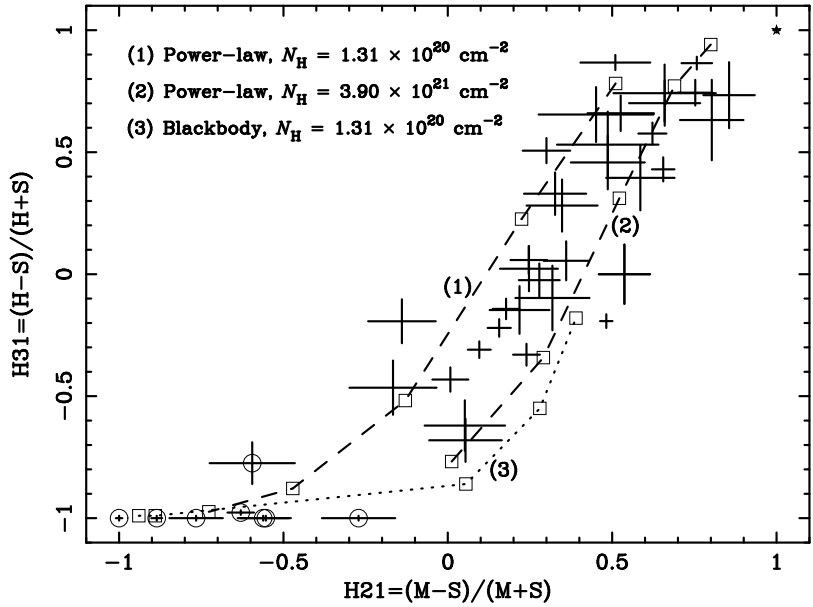}
\caption{Color-color diagram of the resolved point sources. Super soft sources (SSSs) and the hardest source
are marked with open circles and an open star, respectively. Model predictions are shown with dashed and dotted lines. From Top
of the figure downwards, open rectangles on the lines indicate the values of photon indices of $\Gamma=0.0,~1.0,~2.0,~3.0$
and 4.0 for power-law models, and temperature of $kT=0.5,~0.4,~0.3,~0.2$ and 0.1 keV for blackbody model, respectively.}
\end{figure}

\begin{figure}
\centering
\includegraphics[width=11cm,angle=270]{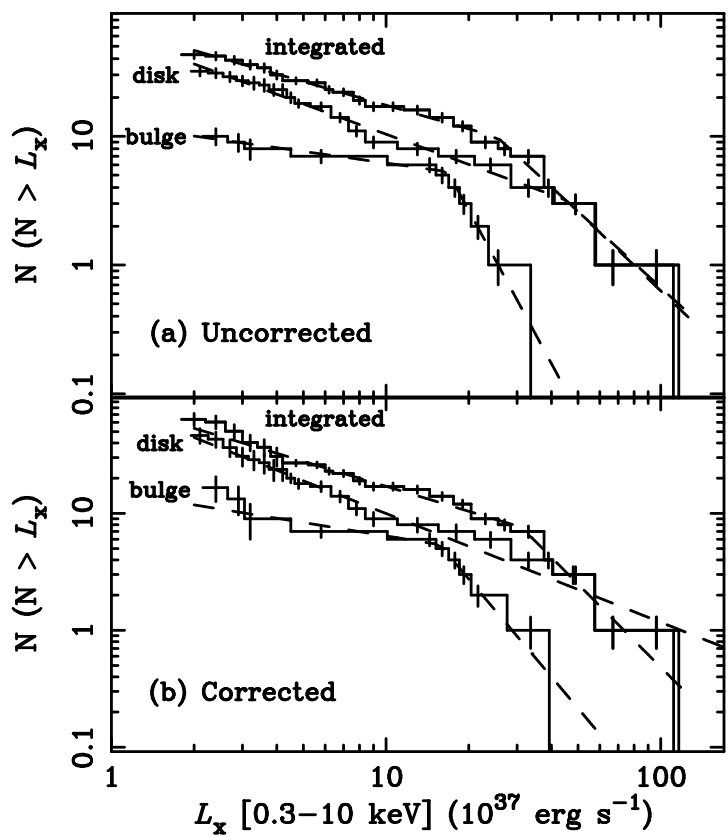}
\caption{XLFs of the detected off-center X-ray point sources uncorrected (a) and corrected (b)
for the incompleteness at the low luminosity end. Models are plotted as dashed lines.
A single power-law model is sufficient to describe the corrected XLF of the disk population, while
in other cases a broken power-law model is needed (\S4.5).}
\end{figure}

\end{document}